\documentclass[twocolumn,amsmath,amssymb,prd,reprint,longbibliography,floatfix,8pt]{revtex4-1}

\usepackage{cancel}
\usepackage{enumitem}
\usepackage{graphicx}
\usepackage{dcolumn}
\usepackage{bm}
\usepackage{hyperref}
\usepackage[switch]{lineno}
\usepackage{float}             
\usepackage{subfigure}
\usepackage{tikz}
\usepackage{hyperref}
\usepackage{tabularx}
\usepackage{adjustbox}
\usepackage{multirow}
\usepackage{xr}
\usetikzlibrary{arrows.meta}

\usepackage{amsmath}
\usepackage{braket}

\begin{document}
\title{Spin-vibronic dynamics in open-shell systems beyond the spin Hamiltonian formalism}
\author{Lorenzo A. Mariano}
\author{Sourav Mondal}
\author{Alessandro Lunghi}
\email{lunghia@tcd.ie}
\affiliation{School of Physics, AMBER and CRANN Institute, Trinity College, Dublin 2, Ireland}

\begin{abstract}
 {\bf Vibronic coupling has a dramatic influence over a large number of molecular processes, ranging from photo-chemistry, to spin relaxation and electronic transport. The simulation of vibronic coupling with multi-reference wave-function methods has been largely applied to organic compounds, and only early efforts are available for open-shell systems such as transition metal and lanthanide complexes. In this work, we derive a numerical strategy to differentiate the molecular electronic Hamiltonian in the context of multi-reference ab initio methods and inclusive of spin-orbit coupling effects. We then provide a formulation of open quantum system dynamics able to predict the time evolution of the electrons' density matrix under the influence of a Markovian phonon bath up to fourth-order perturbation theory. We apply our method to Co(II) and Dy(III) molecular complexes exhibiting long spin relaxation times and successfully validate our strategy against the use of an effective spin Hamiltonian. Our study shed light on the nature of vibronic coupling, the importance of electronic excited states in spin relaxation, and the need for high-level computational chemistry to quantify it.}
\end{abstract}

\maketitle

\section*{Introduction}


The understanding of the interaction among the nuclear degrees of freedom and electronic states, namely the vibronic coupling, is key to understanding the behavior of open-shell coordination compounds and their application in several fields such as photochemistry,\cite{Turro2009-zl} photocatalysis \cite{Wang2015}, ultrafast spectroscopy,\cite{Penfold2018} and molecular magnetism.\cite{Lunghi2023} For instance, vibronic coupling, also called spin-phonon coupling when restricted to a single spin multiple of states, plays a significant role in the thermally induced relaxation of the magnetic moments of single-molecule magnets (SMMs). The latter exhibit long-lived electronic spin states but relaxation processes due to vibronic coupling hinder their potential applications as quantum bits, memory units, and spintronics elements.\cite{Bogani2008,Coronado2019,GaitaArio2019,ZabalaLekuona2021} To date, a successful description of the thermalization process in SMMs has been achieved using an ab initio theory of open quantum systems.\cite{lunghi2022toward,Lunghi2023} This theory allows for the treatment of system-environment interactions on a first principles ground, thus accounting for the coupling between the magnetic ion's spin and molecular vibrations and lattice phonons. Ab initio theory of open quantum systems has already provided important insights into the relaxation dynamics and the resulting loss of spin polarization in SMMs.\cite{Lunghi2017,Goodwin2017,Lunghi2017_2,Moseley2018,Ullah2019,Lunghi2020_2,Reta2021,Blockmon2021,Briganti2021,lunghi2022toward,Mondal2022} 

However, the application of this method has been limited to either systems where a mapping of the lowest electronic states into an effective spin Hamiltonian is possible\cite{lunghi2022toward,Lunghi2023} , or to systems where the electronic structure is well described by single-reference methods.\cite{Park2020,Xu2020,Park2022} Here we aim to remove such restrictions and enable a description of relaxation processes where molecular electronic states are treated explicitly and fully accounting for their multi-reference nature. 

When the spin-Hamiltonian formalism is replaced by a more general approach that uses the full electronic Hamiltonian, the problem can be addressed by adopting the same strategies and concepts employed in the description of excited-state dynamics.\cite{Penfold2018,Matsika2011,Richter2011,Mai2018} In this context, the investigation of ultrafast relaxation mechanisms revolves around a theoretical exploration of the dynamics of the excited states in the limit where the Born-Oppenheimer (BO) approximation breaks down.\cite{Kouppel2007-jn}  A general approach to treat vibronically interacting electronic states would require the use of a diabatic basis where electronic states are no longer parameterized by the nuclear coordinates. Unfortunately, the construction of a diabatic basis for molecular systems is in general not possible\cite{Mead1982}. Nevertheless, when the vibronic coupling matrix elements change slowly with the nuclear coordinates, an effective method to avoid explicitly constructing a diabatic basis is to expand the vibronic coupling matrix around a chosen reference geometry. The linear vibronic coupling (LVC) model, which involves truncation at the first order, has found widespread application in the investigation of medium to large systems.\cite{Zobel2021}

In this work, we apply ab initio open quantum system theory to the study of angular momentum dynamics of two prototypical SMM complexes, i.e.  [Co(C$_3$S$_5$)$_2$](Ph$_4$P)$_2$ \cite{fataftah2014mononuclear} and [Dy(bbpen)Cl]\cite{SMM8}, where H$_2$bbpen= N,N'-bis(2-hydroxybenzyl)-N,N'-bis(2-methylpyridyl)ethylenediamine). Vibronic coupling matrix elements within the LVC approximation are extracted from the full Hamiltonian of the system and used to evaluate the total relaxation time $\tau$ considering both Orbach and Raman relaxation mechanisms. The numerical evaluation of these contributions is discussed and the results are compared with the spin-Hamiltonian approach. We demonstrate that accounting for vibronic coupling among a sufficiently large number of \textit{ab initio} states can result in an alteration of the computed relaxation times compared to the simple treatment of the lowest angular momentum multiplet. Furthermore, we examine the impact of spin-orbit coupling on the simulated dynamics and analyze its effects, thus providing an unprecedently detailed description of spin-phonon relaxation at the quantum mechanical level.

\section*{Theoretical Methods}

{\bf Vibronic Coupling.} We start considering the Molecular Coulomb Hamiltonian (MCH) within the BO approximation which contains the kinetic energy of the electrons ($T_\mathrm{e}$) and the Coulombic interactions electron-electron ($V_\mathrm{e-e}$), electrons-nuclei ($V_\mathrm{e-n}$), and nuclei-nuclei ($V_\mathrm{n-n}$), i.e.  
\begin{equation}\label{eq:MCH}
    \hat{H}^\mathrm{MCH}=T_\mathrm{e}+V_\mathrm{e-e}+V_\mathrm{e-n}+V_\mathrm{n-n}\:.
\end{equation}
 The spin-free MCH wave functions $\ket{\psi^\mathrm{MCH}_i}$ solve the time-independent Schrödinger equation
\begin{equation} 
\hat{H}^\mathrm{MCH}\ket{\psi^\mathrm{MCH}_i}=E_i^\mathrm{MCH}\ket{\psi^\mathrm{MCH}_i}
\label{mch_basis}
\end{equation}
and we assume here that the MCH eigenfunctions are non-degenerate. The MCH Hamiltonian of equation \ref{eq:MCH} alone is not suitable for describing the magnetic properties of SMM systems, for which it is essential to consider the spin-orbit (SO) interaction that allows the mixing of orbital and spin degrees of freedom. Within the one-electron effective approximation\cite{Neese2005,He1996}, the SO Hamiltonian, $\hat{H}_{SO}$, is added to the spin-free MCH one as
\begin{equation} 
    \hat{H}=\hat{H}^\mathrm{MCH}+\hat{H}^\mathrm{SO}\:.
\end{equation}
We will refer to $\mathbf{H}^\mathrm{MCH}_\mathrm{SO}$ as the total Hamiltonian $\hat{H}$ matrix elements in the MCH basis of Eq. \ref{mch_basis}. In general, $\hat{H}_\mathrm{SO}$ couples the spin-free states $\ket{\psi^\mathrm{MCH}_i} $ and $\mathbf{H}^\mathrm{MCH}_\mathrm{SO}$ is non-diagonal in this representation. 

By diagonalizing $\mathbf{H}^\mathrm{MCH}_\mathrm{SO}$, we obtain the SO-corrected eigenvalues, $E_i$, and the corresponding eigensstates, $\ket{\psi_i}$
\begin{equation} 
    \hat{H}\ket{\psi_i}=E_i\ket{\psi_i}\:.
\end{equation}
The matrix $\mathbf{U}$, which contains as columns the coefficients of the SO eigenfunctions with respect to the MCH basis, is such that
\begin{equation}     \mathbf{H}=\mathbf{U}\mathbf{H}^\mathrm{MCH}_\mathrm{SO}\mathbf{U}^{\dagger}\:,
\end{equation}
where $\mathbf{H}$ is the diagonal representation of $\hat{H}$ and
\begin{equation} \label{eq:2}
    \ket{\psi_i}=\sum_{j} U_{i,j}\ket{\psi^\mathrm{MCH}_j}\:.
\end{equation}
Within BO approximation, the nuclear degrees of freedom are kept frozen at their equilibrium geometry $\mathbf{R}_0$ and the interaction between electrons and nuclei is purely electrostatic in nature. This assumption prevents a correct description of all those phenomena in which the nuclear motion happens on a short time-scale leading to significant mixing of nuclear and electronic degrees of freedom. Such mixing is responsible for population transfer between different adiabatic states $\ket{\psi_i}$ and it plays a crucial role in describing relaxation processes in molecular systems.\cite{Daniel2021}\\
We can partly lift the BO approximation considering the LVC expansion in the nuclear displacement of the total Hamiltonian $\hat{H}$
\begin{equation}
    \hat{H}(t)=\hat{H}(\mathbf{R}_0)+\sum_{a} \left. \left (  \frac{\partial \hat{H} }{\partial R_{a}} \right ) \right\vert_{\mathbf{R}_0} R_{a}(t)
    \label{vibronic}
\end{equation}
where the parametric dependence of $\hat{H}$ on the atomic coordinates of the equilibrium geometry $\mathbf{R}_0$ was made explicit. The dependency of $\hat{H}$ on the vibrational degrees of freedom is at the origin of the vibronic coupling between the states $\ket{\psi_i}$ and the nuclei. The choice of employing Cartesian coordinates to evaluate the vibronic coupling matrix elements is primarily driven by practical considerations. However, it is always possible to transform these elements into derivatives with respect to normal modes by using
\begin{equation}
     \frac{\partial }{\partial Q_{{\alpha}}} = \sum_{a}^{3N} \sqrt{\frac{\hbar}{2\omega_{\alpha}m_{a}}} L_{\alpha a}\left( \frac{\partial } {\partial R_a} \right) \:,
    \label{num_diff}
\end{equation}
where $m_i$ is the $i$-th atomic mass and $L_{\alpha i}$ the Hessian matrix eigenvectors.\\

{\bf Electronic dynamics.} The time evolution of the system is determined by the Liouville equation which can be solved within the Born-Markov approximation to obtain the transition rate $W_{ij}$ between two states $\ket{\psi_i}$ and $\ket{\psi_j}$. Here, we describe the nuclei as a phonon bath characterized by normal modes $Q_\alpha$ and associated energies $\hbar\omega_{\alpha}$. Contributions from both one- and two-phonon processes are involved in the relaxation of molecular Kramers systems exhibiting significant magnetic anisotropy. Considering one-phonon processes, the transition rate $\hat{W}^{1-\mathrm{ph}}_{ji}$ reads 
\begin{equation}
    \hat{W}^{1-\mathrm{ph}}_{ji}=\frac{2\pi}{\hbar^2} \sum_{\alpha} | \langle \psi_j| \left(\frac{\partial \hat{H}}{\partial Q_{\alpha}} \right) |\psi_i \rangle  |^2G^{1-\mathrm{ph}}(\omega_{ji}, \omega_{\alpha}) \:, 
    \label{Orbach}
\end{equation}
where $\hbar\omega_{ji}=E_{j}-E_{i}$ and the term $\left( \partial \hat{H} / \partial Q_{\alpha} \right)$ provides the intensity of the vibronic coupling between the electrons and the $\alpha$-phonon $Q_{\alpha}$.  
The function $G^{1-\mathrm{ph}}$ reads
\begin{equation}
G^{1-\mathrm{ph}}(\omega, \omega_{\alpha}) = \delta(\omega-\omega_\alpha)\bar{n}_\alpha +\delta(\omega +\omega_\alpha)(\bar{n}_\alpha +1)    \:,
\label{G1_func}
\end{equation}
where $\bar{n}_\alpha = (e^{ \hbar\omega_{\alpha}/\mathrm{k_B}T} -1)^{-1}$ is the Bose-Einstein distribution accounting for the phonons' thermal population, $\mathrm{k_B}$ is the Boltzmann constant, and the Dirac delta functions enforce energy conservation during the absorption and emission of phonons by the spin system, respectively. The Orbach relaxation mechanism, described by Eq. \ref{Orbach}, considers the transfer of population through the absorption and emission of a single phonon. An alternative pathway for relaxation towards equilibrium is due to two-phonon processes. i.e. the Raman relaxation mechanism. We model two-phonon spin-phonon transitions, $W^{2-\mathrm{ph}}_{ji}$, as
\begin{equation}
    \hat{W}^{2-\mathrm{ph}}_{ji}  =\frac{2\pi}{\hbar^2} \sum_{\alpha\beta}\left | T^{\alpha\beta,+}_{ij} + T^{\beta\alpha,-}_{ij} \right|^2G^{2-\mathrm{ph}} (\omega_{ji}, \omega_{\alpha}, \omega_{\beta})\:,
    \label{Raman}
\end{equation}
where the terms
\begin{equation}
T^{\alpha\beta,\pm}_{ij} = \sum_{k} \frac{ \langle \psi_i| (\partial \hat{H}/\partial Q_{{\alpha}}) |\psi_k\rangle \langle \psi_k| (\partial \hat{H}/\partial Q_{{\beta}})|\psi_j\rangle }{E_k -E_j \pm \hbar\omega_\beta} 
\end{equation}
involve the contribution of all the spin states $|\psi_k\rangle $ at the same time, often referred to as a virtual state. The function $G^{2-\mathrm{ph}}$ fulfills a similar role as $G^{1-\mathrm{ph}}$ for one-phonon processes and includes contributions from the Bose-Einstein distribution and imposes energy conservation. $G^{2-\mathrm{ph}}$ accounts for all two-phonon processes, i.e. absorption of two phonons, emission of two phonons or absorption of one phonon and emission of a second one. The latter process is the one that determines the Raman relaxation rate, and in this case $G^{2-\mathrm{ph}}$ reads
\begin{equation}
G^{2-\mathrm{ph}}(\omega, \omega_{\alpha},\omega_{\beta}) = \delta(\omega-\omega_\alpha+\omega_\beta)\bar{n}_\alpha(\bar{n}_\beta +1)\:.
\label{G2_func}
\end{equation}
Once all the matrix elements $W^{n-\mathrm{ph}}_{ji}$ have been computed, the relaxation time $\tau^{-1}$ can be predicted by simply diagonalizing $W^{n-\mathrm{ph}}_{ji}$ and taking the smallest non-zero eigenvalue. The study of $W^{1-\mathrm{ph}}$ provides the Orbach contribution to the relaxation rate, $\tau^{-1}_\mathrm{Orbach}$, while $W^{2-\mathrm{ph}}$ provides the Raman contribution, $\tau^{-1}_\mathrm{Raman}$. The total relaxation time is thus computed as $\tau^{-1}= \tau^{-1}_\mathrm{Orbach} + \tau^{-1}_\mathrm{Raman}$.\\
The theoretical framework just introduced has been successfully applied in the study of SMMs from first principles.\cite{Mondal2022,lunghi2022toward} In these works, lattice harmonic frequencies $\omega_{\alpha}/2\pi$ and normal modes $Q_{\alpha}$ are computed by finite differentiation at density functional theory (DFT) level of theory, while the Hamiltonian $\hat{H}$ employed to extract the magnetic properties is the spin Hamiltonian (\textit{vide infra}). Although this choice is often well justified when studying SMMs, a general approach to the problem must take into account the full \textit{ab initio} Hamiltonian $\hat{H}$. This choice has the main advantage of preventing any possible loss of information during the construction of the spin Hamiltonian. Moreover, this approach can be applied in situations where it is challenging to determine \textit{a priori} the specific electronic states that contribute to the relaxation process at a given temperature.\\

In this paper, our aim is to explore different levels of theory in the evaluation of $\left( \partial \hat{H} / \partial Q_{\alpha} \right)$ terms and their effect on the calculated relaxation time. The starting point of our analysis is the evaluation of the Hamiltonian derivative $\nabla \hat{H}$ in the \textit{ab initio} wave function basis, i.e. 
\begin{equation} \label{Nabla_H}
\nabla \hat{H}=\sum_{a=1}^{3N}\frac{\partial \hat{H}}{\partial R_a}\:.
\end{equation}
$\nabla \hat{H}$ contains the partial derivative of the Hamiltonian with respect to the $3N$ Cartesian degrees of freedom, where $N$ is the number of atoms in the system. In the next section the details of the evaluation of $\nabla \hat{H}$ are discussed. \\ 

{\bf Hamiltonian derivative.} Vibronic coupling effects are introduced at the first perturbative order in nuclear displacement by considering the gradient of $\hat{H}$, whose matrix elements can be expressed as follows using the Hellmann-Feynman theorem
\begin{equation} \label{Vibronic}
\begin{split}
    & \braket{\psi_j(\mathbf{R}_0)| \nabla \hat{H}|\psi_i(\mathbf{R}_0)}  =\nabla E_i(\mathbf{R}_0) \delta_{ji}\\ & +(E_i(\mathbf{R}_0) -E_j(\mathbf{R}_0))\braket{\psi_j(\mathbf{R}_0)|\nabla \psi_i(\mathbf{R}_0)}\:.
  \end{split}  
\end{equation}
Several theoretical approaches rely on the evaluation of Eq. \ref{Vibronic} to unravel the electronic excited-state dynamics beyond the BO approximation.\cite{Penfold2018} Thus, different electronic structure codes offer the possibility to evaluate forces on atoms and non-adiabatic couplings (NACs) employing both numerical and analytical methods. Nevertheless, the availability of such features is often restricted to specific \textit{ab initio} methods and rarely with the inclusion of SO effects. Hence, our aim is to build a general framework that allows us to evaluate Eq. \ref{Vibronic} numerically, computing the derivative coupling elements $\braket{\psi_j(\mathbf{R}_0)|\nabla \psi_i(\mathbf{R}_0)}$ using the wave function overlap between states at different geometries and with the inclusion of SO effects.
The need to use the wavefunction overlap to evaluate derivative coupling becomes clear by writing the derivative of $\psi_i(\mathbf{R}_0)$ along $R_{a}$ as central finite differentiation
\begin{equation}
    \frac{\partial \ket{\psi_i(\mathbf{R}_0)}}{\partial R_a}= \lim_{\Delta R_{a} \to 0}  \frac{\ket{\psi_i(\mathbf{R}_0 + \Delta R_{a})}- \ket{\psi_i(\mathbf{R}_0 - \Delta R_{a})}}{2\Delta R_{a}}\:.
\end{equation}
The evaluation of $\braket{\psi_j(\mathbf{R}_0)|\nabla \psi_i(\mathbf{R}_0)}$ reduces to the computation of the overlaps $\braket{\psi_j(\mathbf{R}_0)| \psi_i(\tilde{\mathbf{R}})}$, where $\tilde{\mathbf{R}}=\mathbf{R}_0 \pm \Delta R_{a}$ and $a$ denotes one of the $3N$ displacement direction used in the numerical differentiation.
For this last purpose, it is more practical to use Eq. \ref{eq:2} and work with the MCH wave functions since their composition in terms of Slater determinant coefficients, molecular orbitals (MOs), and atomic orbitals (AOs) can be obtained directly from \textit{ab initio} quantum chemistry calculations. The derivative coupling between the SO states $i$ and $j$ can be rewritten as
\begin{equation}
   \braket{\psi_j(\mathbf{R}_0)|\nabla \psi_i(\mathbf{R}_0)} = K^\mathrm{SO}_{ij}(\mathbf{R}_0)+K^\mathrm{U}_{ij}(\mathbf{R}_0)
 \label{Vibronic_dec}
\end{equation}
where $K^\mathrm{SO}_{ij}(\mathbf{R}_0)$ and $K^\mathrm{U}_{ij}(\mathbf{R}_0)$, obtained using the product rule of differentiation, represent the derivative couplings between MCH states in the SO basis and the variation of the rotational matrix $\mathbf{U}$ respectively, i.e.
\begin{align}\label{eq:KSO}
    &K^\mathrm{SO}_{ij}(\mathbf{R}_0)= \\
    &\sum_{k l} U^*_{k i} (\mathbf{R}_0)\braket{\psi^{MCH}_k(\mathbf{R}_0)|\nabla \psi^{MCH}_l(\mathbf{R}_0)} U_{l j}(\mathbf{R}_0)
\end{align}
and
\begin{equation} \label{KU}
    K^\mathrm{U}_{ij}(\mathbf{R}_0)=\sum_{k } U^*_{k i}(\mathbf{R}_0) \nabla  U_{k j}(\mathbf{R}_0)\:.
\end{equation}
It is important to notice that in order to evaluate Eq. \ref{Vibronic} special attention must be paid to the calculation of the terms $K^\mathrm{SO}_{ij}(\mathbf{R}_0)$ and $K^\mathrm{U}_{ij}(\mathbf{R}_0)$ to avoid phase inconsistency between wave functions calculated at different geometries.\\ \\
{\bf Phase correction.} The electronic MCH wavefunctions are obtained by solving the eigenvalue problem \ref{eq:MCH} parametrized with the nuclear coordinates of the system. Consequently, if $\ket{\psi^\mathrm{MCH}_i(\mathbf{R})}$ is a valid solution to the problem for a given geometry $\mathbf{R}$ so is $e^{i\phi_i(\mathbf{R})}\ket{\psi^\mathrm{MCH}_i(\mathbf{R})}$. The real number $\phi_i(\mathbf{R})$ ultimately depends on the specific implementation of the eigensolver. This arbitrariness of the phase factor value must be taken into account when calculating the overlap terms $\braket{\psi^\mathrm{MCH}_\alpha(\mathbf{R}_0)| \psi^\mathrm{MCH}_\beta(\tilde{\mathbf{R}})}$ that enter Eq. \ref{eq:KSO}. \cite{Akimov2018,Zhou2019}
In order to fix the phase, the wave functions at displaced geometries are transformed as $ \ket{\psi^\mathrm{MCH}_i(\tilde{\mathbf{R}})} \rightarrow f^*_i \ket{\psi^\mathrm{MCH}_i(\tilde{\mathbf{R}})}$.  The phase-correction factor $f_i$ is defined as 
\begin{equation}
    f_i=\frac{\braket{\psi^\mathrm{MCH}_i(\mathbf{R}_0)| \psi^\mathrm{MCH}_i(\tilde{\mathbf{R}})}}{|\braket{\psi^\mathrm{MCH}_i(\mathbf{R}_0)| \psi^\mathrm{MCH}_i(\tilde{\mathbf{R}})}|}\:,
\end{equation}
and it is chosen such that the diagonal elements of the overlap matrix are real and positive.
The vector $\Vec{f}$ then contains the values of the phase-correction factors $f_i$ for each of the MCH wavefunction $ \ket{\psi^\mathrm{MCH}_i(\tilde{\mathbf{R}})}$ as elements, and it is used to transform all the phase-dependent matrices expressed in the MCH basis, such as $\mathbf{H}^\mathrm{MCH}_\mathrm{SO}$ and the angular momentum operators $\mathbf{J}^\mathrm{MCH}_i$, $\mathbf{L}^\mathrm{MCH}_i$, and $\mathbf{S}^\mathrm{MCH}_i$ ($i=x,y,z$).\\
A further phase-correction step is required when $\mathbf{H}^\mathrm{MCH}_\mathrm{SO}$ is diagonalized to obtain the SO eigenfunctions. More precisely, we are interested in obtaining phase consistency between the rotational matrices $\mathbf{U}$ at displaced coordinates as these enter the expression of $\nabla \hat{H}$ through $\mathbf{K}^\mathrm{U}$. When dealing with Kramers systems, the degeneracy of the SO states has also to be considered in addition to the arbitrary phase introduced by solving the eigenvalue problem. The phase tracking algorithm proposed by S. Mai et al.\cite{Mai2015} is used to correct the phase of the rotational matrix $\mathbf{U}(\tilde{\mathbf{R}})$ with respect to $\mathbf{U}(\mathbf{R}_0)$. Firstly, the overlap matrix $\mathbf{U}(\mathbf{R}_0)^\dagger \mathbf{U}(\tilde{\mathbf{R}})$ is computed and put into block-diagonal form by setting to zero all the matrix elements belonging to non-degenerate eigenstates. We refer to such a matrix as $\mathbf{M}$ and, since the eigenvalues of $\mathbf{H}^\mathrm{MCH}_\mathrm{SO}$ correspond to the energies of the Kramers doublets, each diagonal block is two dimensional. Subsequently, $\mathbf{M}$ is Löwdin orthonormalized to extract the matrix $\mathbf{\Phi}$ such that the transformation  $\mathbf{U}(\tilde{\mathbf{R}}) \rightarrow \mathbf{U}(\tilde{\mathbf{R}}) \mathbf{\Phi}^\dagger$ gives rotational matrices with the correct phase factors. In our implementation, $\mathbf{\Phi}$ is obtained by computing the single value decomposition (SVD) of $\mathbf{M}$, i.e.
\begin{equation}
    \mathbf{M}=\mathbf{u} \mathbf{\sigma} \mathbf{v}^*
\end{equation}
 and 
 \begin{equation}
    \mathbf{\Phi}=\mathbf{u}\mathbf{v}^* \:.
\end{equation}

{\bf Spin Hamiltonian.} The most common way of approaching the problem of calculating relaxation times in SMMs is to go through the spin-Hamiltonian formalism that, by incorporating the relevant degrees of freedom and interactions, provides a simplified description that can effectively reproduce experimental observations.\cite{Maurice2009,Jung2019,Chibotaru2012,Ungur2017,Riedl2019} Within this theoretical framework, it is possible to evaluate Eqs. \eqref{Orbach}-\eqref{G2_func} in a reduced Hilbert space spanned by the ground $J$-multiplet of the bare magnetic ion of the system, where $J$ denotes the total angular momentum quantum number.\cite{Mondal2022,lunghi2022toward} The mapping of the total Hamiltonian $\hat{H}$ into the model space gives the generalised spin Hamiltonian $\hat{H}_s$ which can be expressed as 
\begin{equation}
    \hat{H}_{s}= \sum_{l=2\: (\text{even})}^{2J} \sum_{m=-l}^{l} B_{m}^{l} \hat{O}_{m}^{l}\:,
    \label{CFHam}
\end{equation}
  where the operators $\hat{O}_{m}^{l}$ are a tesseral function of the total angular momentum operator $\hat{J}$ of rank $l$ and order $m$. To obtain the spin Hamiltonian coefficients from first principles, the lowest $2J+1$ \textit{ab initio} wavefunctions are put in a one-to-one correspondence with the magnetic ground state of the ion. To do so, the operator $\hat{J}_z$ in the basis of the lowest $2J+1$ \textit{ab initio} wavefunctions is diagonalized to obtain the basis states of the spin Hamiltonian $\ket{\tilde{J},\tilde{m}_j}$. Note that this basis corresponds to Zeeman states $ \ket{J,m_j} $ ($-J<m_j<J$) only if the $2J+1$ lowest \textit{ab initio} wavefunctions are completely decoupled from the other excited states, i.e. only if the operator $\hat{J}_z$ in the \textit{ab initio} basis is block diagonal. Thus, in practical computations, it is crucial to verify whether the eigenvalues of $\hat{J}_z$ in the initial \textit{ab initio} basis closely resemble the expected values that would be derived for purely Zeeman states.  If this correspondence holds true, the \textit{ab initio} Hamiltonian is expressed in this new basis and rewritten using Steven's operators.\cite{VandenHeuvel2013} Finally the set of coefficients $B_{m}^{l}$ are adjusted to reproduce the Hamiltonian matrix elements of the lowest $2J+1$ \textit{ab initio} states
  \begin{equation} \label{mapping} 
  \bra{\tilde{J},\tilde{m}_j}\hat{H}_{s}\ket{\tilde{J},\tilde{m}_{j'}}=\bra{\tilde{J},\tilde{m}_j}\hat{H}\ket{\tilde{J},\tilde{m}_{j'}}\:.
 \end{equation}

  In a similar way, we can define the operator 
  \begin{equation}
    \nabla \hat{H}_{s}= \sum_{a}^{3N} \sum_{l=0}^{2J} \sum_{m=-l}^{l}  \left( \frac{\partial B_{m}^{l}}{\partial R_{a}} \right) \hat{O}_{m}^{l}\:,
    \label{GradCFHam}
\end{equation}
which describes the coupling of spin and atomic displacements. We will refer to the elements $\partial B_{m}^{l} / \partial R_{a}$ as spin-phonon coupling coefficients and to $\nabla \hat{H}$ as vibronic coupling operator. 
The spin-phonon coupling coefficients can computed in two different ways, either by numerically differentiating the parameters $B_{m}^{l}$, or by projecting the vibronic coupling operator onto the lowest $2J+1$ ab initio states similarly to the static case,\cite{Staab2022}
  \begin{equation} \label{mapping_grad} 
  \bra{\tilde{J},\tilde{m}_j}\nabla \hat{H}_{s}\ket{\tilde{J},\tilde{m}_{j'}}=\bra{\tilde{J},\tilde{m}_j}\nabla \hat{H}\ket{\tilde{J},\tilde{m}_{j'}}\:.
 \end{equation}

\section*{Computational Methods}
\begin{figure}[h!]
    \centering
    \includegraphics[scale=0.15]{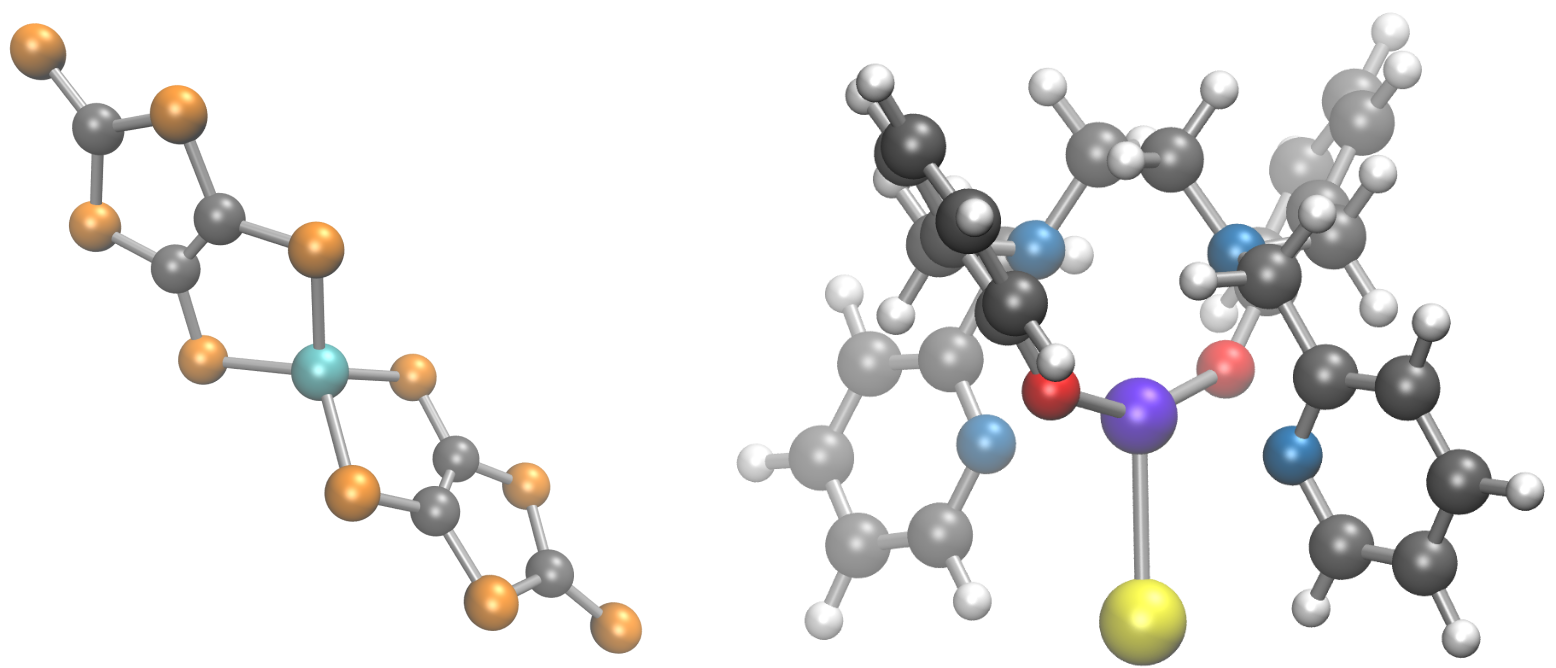}
    \caption{Molecular complexes studied in this work. Left: [Co(C$_3$S$_5$)$_2$](Ph$_4$P)$_2$. Right: [Dy(bbpen)Cl]. Color code: cyan for cobalt, purple for dysprosium, blue for nitrogen, red for oxygen, orange for sulfur, yellow for chlorine, grey for carbon, and white for hydrogen.}
    \label{image1}
    \hfill
\end{figure}

Two complexes were chosen to conduct this study: (\textbf{1}) [Co(C$_3$S$_5$)$_2$](Ph$_4$P)$_2$ and (\textbf{2}) [Dy(bbpen)Cl], where H$_2$bbpen= N,N'-bis(2-hydroxybenzyl)-N,N'-bis(2-methylpyridyl)ethylenediamine). Their optimized molecular structures are shown in Figure \ref{image1}. These compounds serve us as case a study for SMMs based on Co(II) and Dy(III) ions and with long relaxation time.\cite{ZabalaLekuona2021}
Mondal et al. have previously conducted cell and geometry optimization as well as simulations of $\Gamma$-point phonons for the two compounds under investigation.\cite{Mondal2022} They utilized the Perdew-Burke-Ernzerhof (PBE) functional along with DFT-D3 dispersion corrections.\cite{perdew1996generalized,grimme2010consistent} In the present work, we have reused the same dataset.\\ 

ORCA\cite{neese2020orca} had been used to compute molecular electronic properties using the state-averaged (SA) complete active space self-consistent field (CASSCF) method. For \textbf{1}, the active space was built with seven electrons in the five 3$d$ orbitals and all states with multiplicity six were considered. For \textbf{2}, an active space of nine electrons in the seven 4$f$ orbitals was considered and all states with multiplicity six were considered. The RIJCOSX approximation for the Coulomb and Exchange integral was used for both systems. The basis sets DKH-def2-QZVPP for Co atoms, DKH-def2-SVP for H and SARC2-DKH-QZVP for Dy atoms were used. DKH-def2-TZVPP basis set has been used for the rest of the atoms present in the systems. \\

An in-house Python code has been developed to evaluate $\nabla \hat{H}$. This program reads AOs overlap, MOs, and the CI composition of the MCH wavefunctions from ORCA's output files and generates a valid input file for the program WFOVERLAP.\cite{Plasser2016,Plasser2016_2} The latter is subsequently used to compute the overlap matrix between MCH states at different geometries needed to compute the non-adiabatic couplings $\braket{\psi_i^\mathrm{MCH}|\nabla \psi_j^\mathrm{MCH}}$. Our program allows to evaluate $\nabla \hat{H}$ in the SO basis while correcting the phase of the electronic wavefunctions and of the rotational matrices. The numerical differentiation is performed using central differentiation around the equilibrium geometry with a step of $0.01$ \AA.\\

Second- and fourth-order time-dependent perturbation theories have been used to simulate one- and two-phonon processes, respectively. The software MolForge is used for these simulations and it is freely available at github.com/LunghiGroup/MolForge\cite{lunghi2022toward}. As discussed elsewhere, the simulation of Kramers systems in zero external field requires the use of the non-diagonal secular approximation, where population and coherence terms of the density matrix are not independent of one another. This is achieved by simulating the dynamics of the entire density matrix for one-phonon processes\cite{lunghi2019phonons,lunghi2022toward}. An equation that accounts for the dynamics of the entire density matrix under the effect of two-phonon processes resulting from fourth-order time-dependent perturbation theory is not yet available. However, it is possible to remove the coupling between population and coherence terms by applying a small magnetic field along the magnetic easy-axis to break Kramers degeneracy\cite{lunghi2022toward}. Here we employ the latter strategy to simulate Raman relaxation.

\section*{Numerical results} 

\begin{figure}[h!]
    \centering
    \includegraphics[scale=0.25]{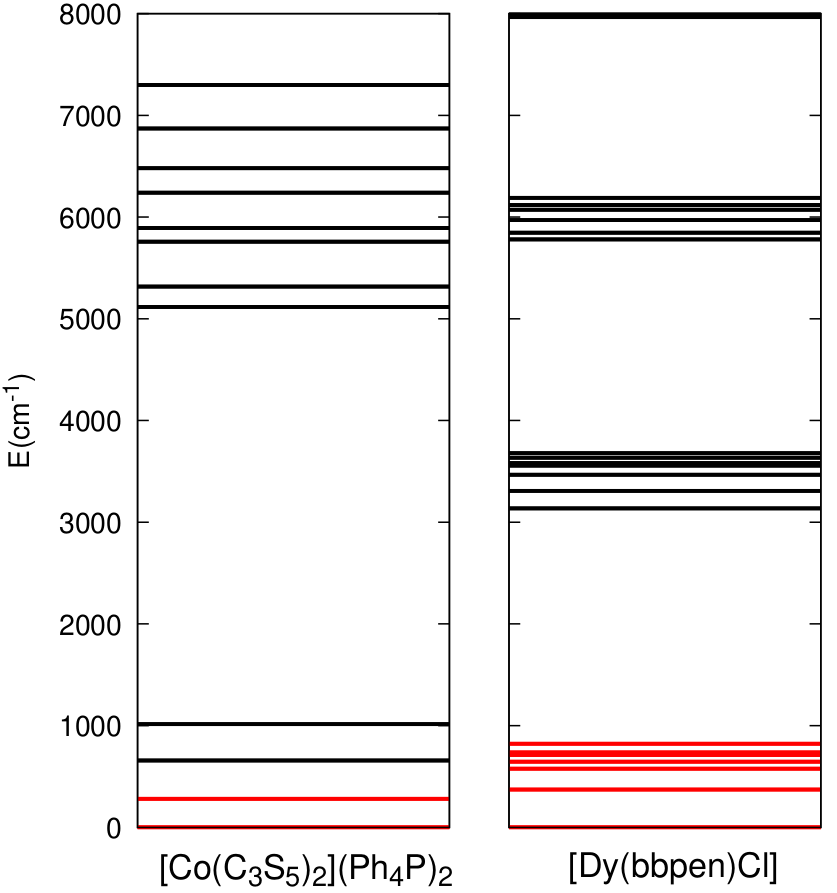}
    \caption{Energy eigenvalues in cm$^{-1}$ computed for $\mathbf{1}$ (left) and $\mathbf{2}$ (right) up to 8000 cm$^{-1}$. The eigenvalues corresponding to the lowest $2J+1$ states are highlighted in red. The scale's zero point is aligned with the energy of the first Kramers doublet.}
    \label{spectra}
    \hfill
\end{figure}

{\bf Electronic structure and spin-phonon coupling.} The computed energy spectra of the two compounds under investigation are reported in Figure \ref{spectra}. The energies of the lowest $2J+1$ states used to build the spin Hamiltonian span 280.2 cm$^{-1}$ for $\mathbf{1}$ and 821 cm$^{-1}$ for $\mathbf{2}$. It is noteworthy that the energy gap between the ground state $J$ multiplet and the next first excited state is significantly smaller in the case of complex $\mathbf{1}$ compared to complex $\mathbf{2}$. In the former case, such energy separation amounts to 376.4 cm$^{-1}$ and is comparable to the energy spacing between the first two Kramer doublets. On the other hand, for complex $\mathbf{2}$, the lowest $2J+1$ states are well separated from the first next excited state by an energy of 2424.5 cm$^{-1}$. \\

\begin{figure*}[htbp!]
    \centering
    \includegraphics[scale=0.25]{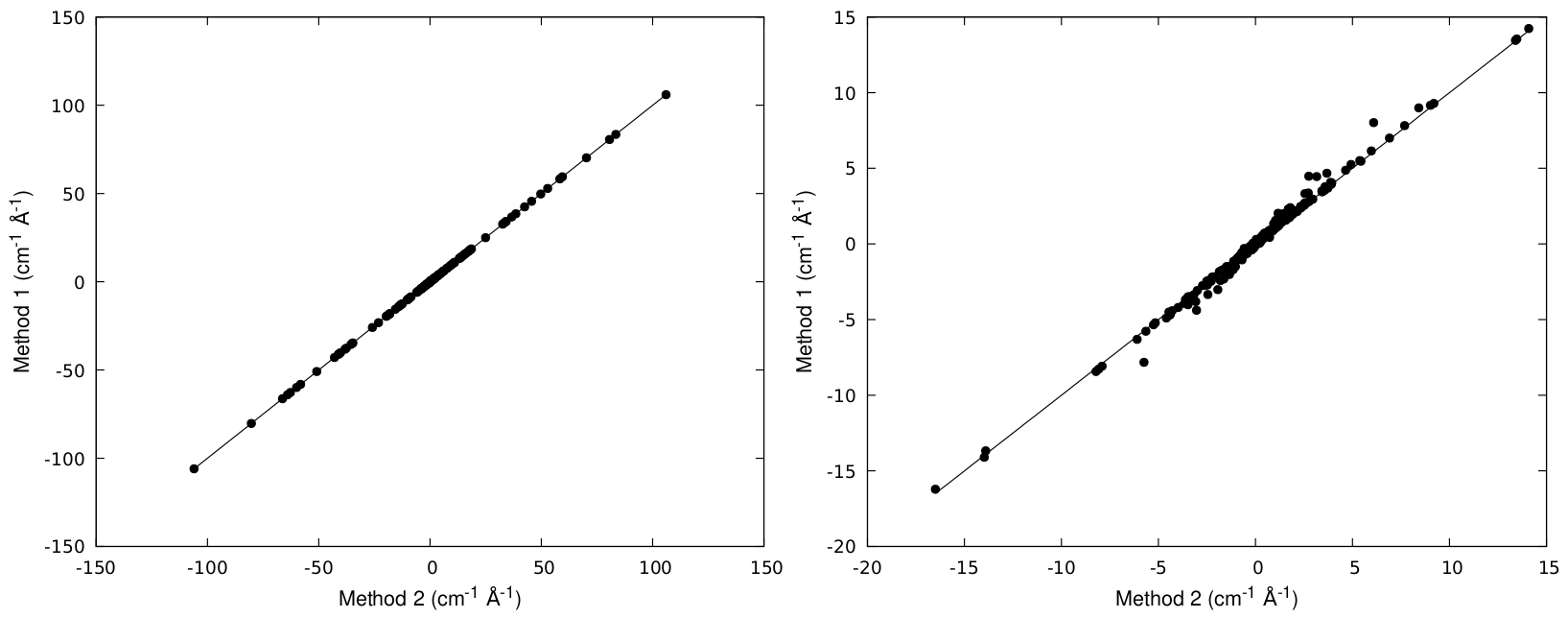}
    \caption{Comparison between the computed matrix elements $\partial B_{m}^{l} / \partial R_{a}$ obtained by numerical differentiation of $B_{m}^{l}$ (Method 1) and by projecting NACs between ab initio wavefunctions into the spin Hamiltonian subspace (Method 2). Left panel shows results for complex $\mathbf{1}$ and right panel for complex $\mathbf{2}$.} 
    \label{B_comparison}
    \hfill
\end{figure*}

As a first step in our analysis, we benchmark the quality of the computed $\nabla\hat{H}$ matrix elements by comparing the derivative of the SH coefficients $\partial B_{m}^{l} / \partial R_{a}$ obtained i) by differentiating $B_{m}^{l}$ from equation \ref{CFHam} and ii) by using directly equation \ref{mapping_grad}. For the latter, the \textit{ab initio} eigenfunctions are written in terms of (pseudo) angular momentum states as
\begin{equation}
    \ket{\psi_i}=\sum_\alpha \mathcal{W}_{i,\alpha} \ket{\tilde{J},\tilde{m}_\alpha} \:,
    \label{abinitio_J}
\end{equation}
and used to write the total Hamiltonian in the basis of the lowest $2J+1$ eigenfunctions as 
\begin{equation}
    \hat{H}=\sum^{2J+1}_n E_n\ket{\psi_n}\bra{\psi_n} \:.
\end{equation}
Finally, vibronic coupling matrix elements are computed using equations \ref{Vibronic},  \ref{eq:KSO}, and \ref{KU}. Because of the definition \ref{abinitio_J}, the terms $K^\mathrm{SO}_{ij}(\mathbf{R}_0)$ and $K^\mathrm{U}_{ij}(\mathbf{R}_0)$ are now expressed with respect to the $J$-multiplet $\ket{\tilde{J},\tilde{m}_\alpha}$ and the rotation matrix $\bm{\mathcal{W}}(\mathbf{R}_0)$. Since the Zeeman states $\ket{\tilde{J},\tilde{m}_\alpha}$ contain only angular momentum degrees of freedom, the terms $K^\mathrm{SO}_{ij}(\mathbf{R}_0)$ vanish and the vibronic coupling only depends on the energies $E_n$ of the \textit{ab initio} eigenfunctions and on the rotation matrix $\bm{\mathcal{W}}(\mathbf{R}_0)$. The results are shown in Figure \ref{B_comparison} where the coefficients $\partial B_{m}^{l} / \partial R_{a}$ computed with the two approaches are compared. As a measure of the linear relationship between the two sets of values, we calculate the Pearson's Product Moment Correlation Coefficient (PPMCC) and the Root Mean Square Error (RMSE). For complex $\mathbf{1}$ the PPMCC and the RMSE are 0.9999 and 0.008 cm$^{-1}$ \AA$^{-1}$ respectively, while for $\mathbf{2}$ the computed PPMCC is 0.9958 and the RMSE is 0.036 cm$^{-1}$ \AA$^{-1}$.

\begin{figure*}[htbp!]
    \centering
    \includegraphics[scale=0.25]{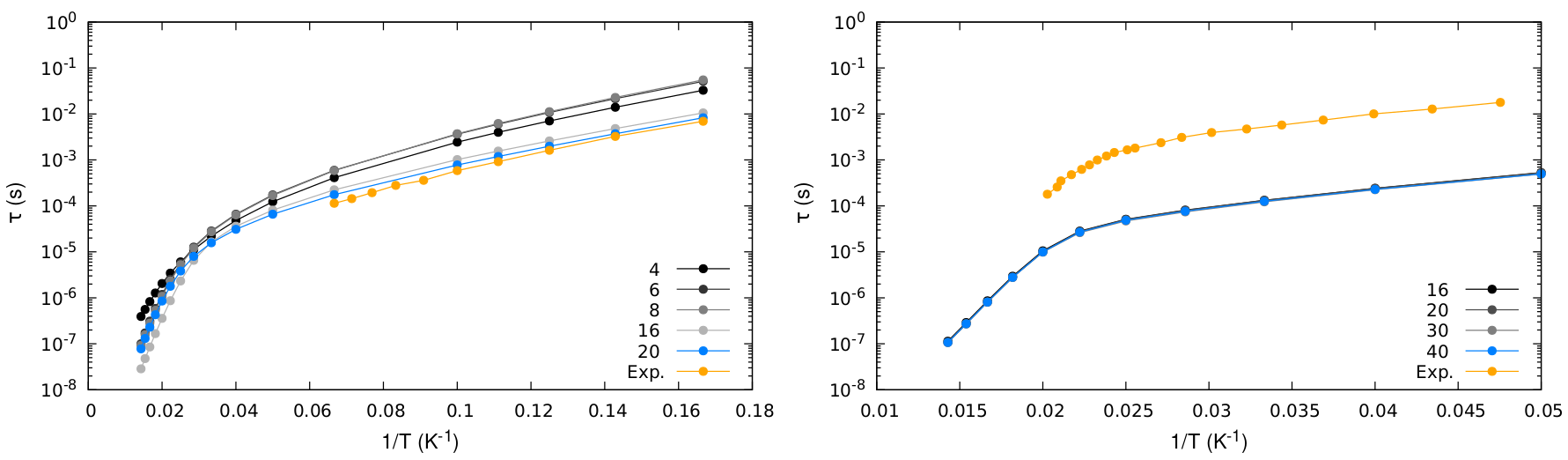}
    \caption{Raman Spin-phonon relaxation times $\tau$ as a function of temperature for different sizes of the Hilbert space used to compute the two-phonon transition rate $\hat{W}^{2-ph}$. In orange experimentally extracted data are reported. The left panel shows results for complex $\mathbf{1}$ and the right panel for complex $\mathbf{2}$.}
    \label{H_convergence}
    \hfill
\end{figure*}

{\bf Relaxation time with the full Hilbert space.} Analysis of the electron dynamics within ab initio open quantum system theory was conducted to compare the computed relaxation times $\tau$ obtained through both the proposed full Hamiltonian method and within the spin-Hamiltonian theory. Under the full Hamiltonian framework, the computation of matrix elements $\hat{W}^{1-\mathrm{ph}}$ and $\hat{W}^{2-\mathrm{ph}}$ is not confined to the subset of the Hilbert space defined by the ground state $J$-multiplet and Figure \ref{H_convergence} shows Raman relaxation obtained using various Hilbert space sizes. We do not display relaxation times associated with the Orbach mechanism because they remain unchanged when expanding the dimension of the Hilbert space due to the lack of resonant phonons with the high-energy excited states introduced. In the case of complex $\mathbf{2}$, convergence is achieved immediately, and the Raman relaxation times remain consistent with those obtained using a minimal $2J+1$-dimensional Hilbert space. However, a different behavior is observed for complex $\mathbf{1}$, where convergence is only achieved when utilizing Hilbert spaces with more than twenty states. This difference is consistent with the spectra reported in Figure \ref{spectra}, where for complex $\mathbf{1}$, the energy of the first excited state outside the ground-state $J$-multiplet and the density of states is significantly lower compared to compound $\mathbf{2}$. \\
\begin{figure*}[htbp!]
    \centering
    \includegraphics[scale=0.25]{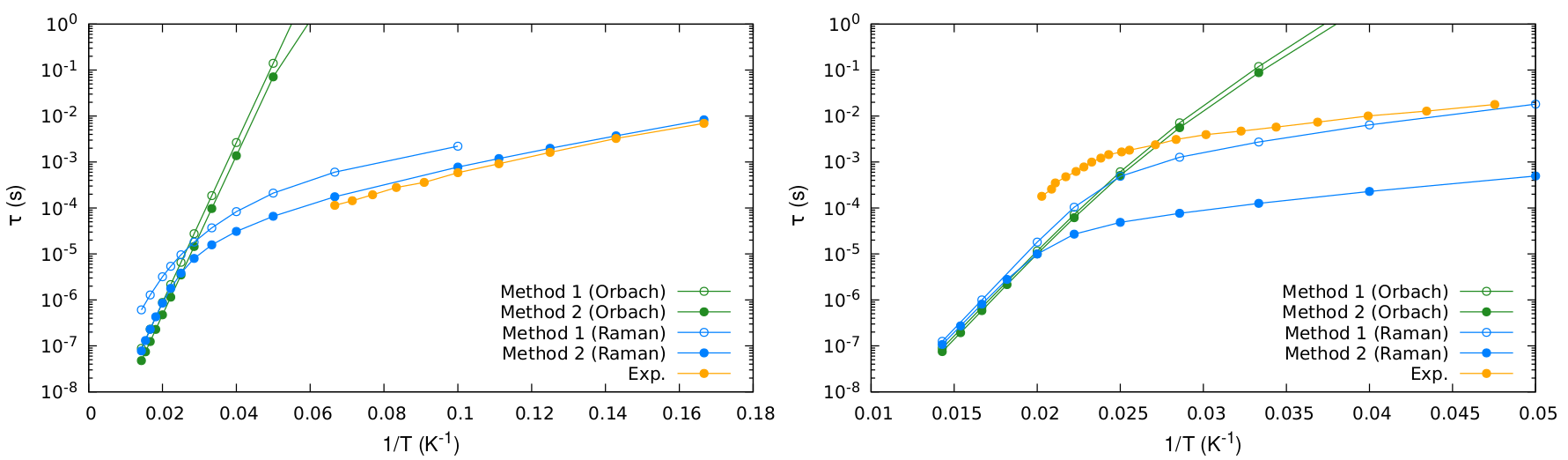}
    \caption{Orbach (green) and Raman (blue) relaxation times $\tau$ as a function of temperature computed within the spin-Hamiltonian framework (Method 1, empty dots) and by employing the full Hamiltonian approach (Method 2, filled dots).  In orange experimentally extracted date are reported. Left panel shows results for complex $\mathbf{1}$ and right panel for complex $\mathbf{2}$.}
    \label{Relax_times}
    \hfill
\end{figure*}
\begin{figure*}[htbp!]
    \centering
    \includegraphics[scale=0.25]{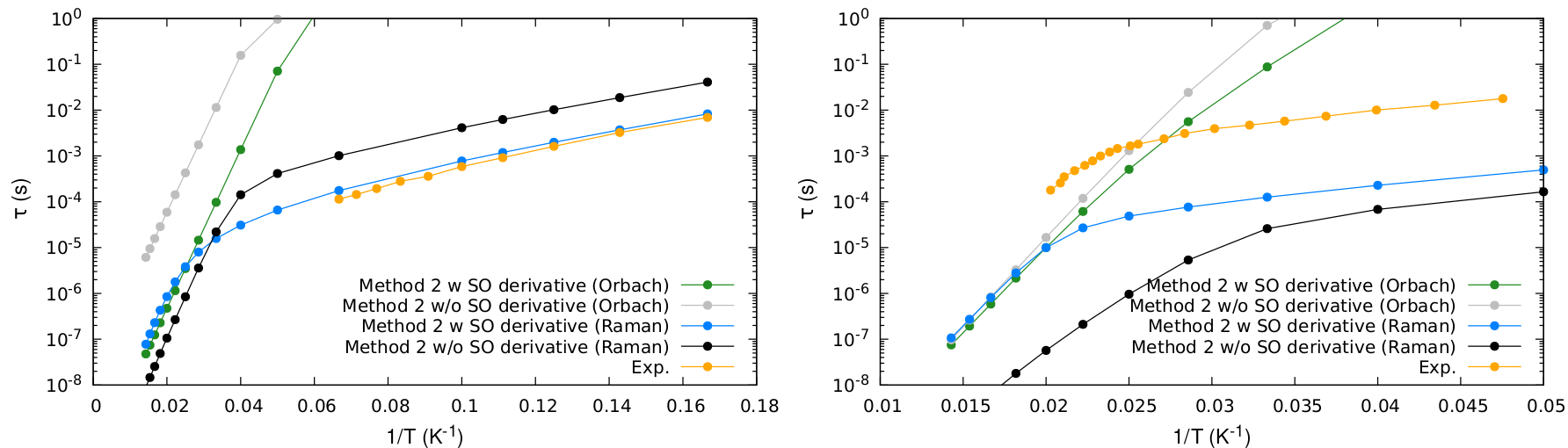}
    \caption{Orbach (green, gray) and Raman (blue, black) relaxation times $\tau$ as a function of temperature computed using the full Hamiltonian approach with and without the inclusion of SO derivatives.  In orange experimentally extracted date are reported. Left panel shows results for complex $\mathbf{1}$ and right panel for complex $\mathbf{2}$.}
    \label{Relax_no_soc}
    \hfill
\end{figure*}

{\bf Comparison with the spin Hamiltonian.} Next, we compare the converged relaxation time values obtained within the full Hamiltonian framework to those obtained using the spin-Hamiltonian approximation. Results are reported in Figure \ref{Relax_times}. The decomposition of the computed $\tau$ in terms of Orbach and Raman contributions and experimental extracted values of $\tau$ are also reported. At elevated temperatures, the dominance of Orbach relaxation is observed, whereas the influence of the Raman mechanism emerges only at lower temperatures. This behavior is mainly due to the temperature dependency of the phonon population. Orbach processes involve the absorption of a single phonon in resonance with the excited electronic states, if resonant phonons are thermally activated then the Orbach mechanism dominates the relaxation process. As temperature decreases, resonant phonons are no longer populated and relaxation happens through absorption/emission of a pair of out-of-resonance low-energy phonons, i.e. the Raman mechanism.\\

The Orbach relaxation times, determined using both methodologies, generally display similar orders of magnitude, with the most significant deviation being a factor of two observed for compound $\mathbf{1}$. In contrast, substantial differences emerge when comparing the computed Raman relaxation times for both complexes. In the case of $\mathbf{1}$ the relaxation time obtained using the converged Hilbert space is three times shorter than the one predicted with the use of the spin Hamiltonian over the entire temperature range. This phenomenon is expected due to the growing number of excited states considered in the spin dynamics, which can facilitate coupling between the ground-state Kramers doublet $m_j=\pm J$, even at relatively low temperatures. Overall, better agreement with experimental results is achieved when utilizing the full Hamiltonian framework. 
Turning our attention to complex $\mathbf{2}$, significant differences between the two methods emerge in the Raman relaxation mechanism at low temperatures. Within the full Hamiltonian approach, the relaxation times are more than an order of magnitude smaller compared to both experimental results and the outcomes obtained using the spin-Hamiltonian framework. This substantial deviation persists even when considering only the lowest $2J+1$ ab initio states. One could be tempted to blame this deviation on some technical aspects of the calculations, primarily the displacement step utilized in the evaluation of the non-adiabatic coupling matrix elements. However, a strong correlation exists when comparing the coefficients $\partial B_{m}^{l} / \partial R_{a}$ obtained with the two methodologies (see. Figure \ref{B_comparison}). This leaves us with two options: i) the process of constructing a spin Hamiltonian is less susceptible to numerical noise or ii) the spin Hamiltonian is not in this case able to fully capture the physics behind spin-vibronic coupling. \\

{\bf Spin-orbit coupling effect.} So far, SO coupling and its dependency on the atomic coordinates have been fully accounted for through the evaluation of the terms $K^\mathrm{SO}_{ij}(\mathbf{R}_0)$ and $K^\mathrm{U}_{ij}(\mathbf{R}_0)$. Here, our intent is to show how considering the variation of the SO operator with respect to nuclear coordinates in the evaluation of vibronic coupling matrix elements can affect the computed relaxation times $\tau$. To achieve this, we assume the term $K^\mathrm{U}_{ij}(\mathbf{R}_0)$ to be negligible in Eq. \ref{Vibronic_dec} and proceed to reevaluate $\hat{W}^{1-\mathrm{ph}}$ and $\hat{W}^{2-\mathrm{ph}}$. This approximation is equivalent to assuming that the SO coupling matrix elements remain constant at their equilibrium values when a small displacement in atomic coordinates is applied. In Figure \ref{Relax_no_soc}, we show Orbach and Raman relaxation times calculated within the full Hamiltonian framework, both with and without the inclusion of SO derivatives, in comparison to experimental results. For both complexes $\mathbf{1}$ and $\mathbf{2}$, we observe that Orbach relaxation times are slower when the matrix elements $K^\mathrm{SO}_{ij}(\mathbf{R}_0)$ are excluded from the calculation. In the case of $\mathbf{1}$, omitting the SO derivative leads to Orbach relaxation times that are over two orders of magnitude larger. However, for compound $\mathbf{2}$, the effect of the SO derivative only marginally impacts the Orbach relaxation times in the high-temperature regime. Regarding the Raman relaxation mechanism, we note distinct behavior for the two systems under investigation. In $\mathbf{1}$, removing the $K^\mathrm{SO}_{ij}(\mathbf{R}_0)$ matrix elements results in relaxation times that are up to three times larger compared to the full Hamiltonian picture. Conversely, in complex $\mathbf{2}$, the Raman relaxation times are consistently smaller when the SO derivative is excluded, with a difference of one order of magnitude in the low-temperature regime. Furthermore, in all cases, significant changes between Raman and Orbach relaxation times are observed in the high-temperature limit.

\section*{Discussion}

The presented numerical method allows the prediction of Orbach and Raman relaxation times in SMMs using \textit{ab initio} wavefunctions from electronic structure methods avoiding the construction of a spin Hamiltonian, thus generalizing previously proposed strategies \cite{Lunghi2023}. \\

Overall, the results are in good agreement with experiments, but interestingly the inclusion of the full electronic Hilbert space improves over spin Hamiltonian results for [Co(C$_3$S$_5$)$_2$]$^{-}$ but worsens them for [Dy(bbpen)Cl]. We believe that multiple effects are at play. On the one hand, the results for [Co(C$_3$S$_5$)$_2$]$^{-}$, clearly show that multiple electronic excited states can contribute to the Raman relaxation, going beyond the ground-state spin multiplet. This is only possible with the proposed approach and it is an effect that must be fully accounted for going forward. On the other hand, the use of a full Hilbert space leads to the largest deviation with respect to experiments in [Dy(bbpen)Cl] despite being the most accurate method in principles. This points to a potential presence of error cancellation effects in the use of the spin Hamiltonian in [Dy(bbpen)Cl] and highlights the need for extra care in comparing simulations and experiments. In general, discrepancies between theory and experiments up to one order of magnitude are not unprecedented for this kind of simulation. These differences might be in part ascribed to the presence of relaxation mechanisms that extend beyond the scope of this study, particularly the absence of spin-spin dipolar cross-relaxation.\cite{Srivastava1980,Bloembergen1948} Fluctuations in the computed relaxation times can also be attributed to the specific limitations of the ab initio methods. In this regard, three main points deserve further analysis: i) phonon calculations are limited to the $\Gamma$-point, neglecting any effects arising from acoustic phonons and the dispersion of optical modes. As shown previously, this effect might lead to substantial deviations at low temperature; ii) whilst the CASSCF electronic structure method is able to capture the multiconfigurational nature of SMMs, it is not efficient in accounting for dynamical correlation effects, which could be significant in this context; iii) the evaluation of wavefunction overlap through finite differences is a general and powerful approach, but a careful convergence of NAC vectors with respect to the differentiation step size is required.\\

Going beyond methodological considerations, the results presented here shed light on the origin of spin-phonon coupling at the quantum mechanical level. For instance, the proposed method is able to isolate the key role of SO coupling derivatives in the study of spin-phonon dynamics in SMMs. To the best of our knowledge, the importance of this term has not been considered in previous studies,\cite{Staab2022}. Moreover, the study of the Dy compound shows evidence that the spin Hamiltonian approximation might not always be fully justified and that the full wavefunction contains additional information. We envision that the extension of our method to other molecular complexes will further help unravel the contributions to spin relaxation of pure electronic origin. The present study also paves the way for a systematic exploration of the role of electronic excited-state dynamics in open-shell transition metal and lanthanide-based coordination compounds and its interplay with spin dynamics.\cite{Gorgon2023,Bayliss2020,Frster2020} Indeed, effects such as inter-system crossing and internal convention also find their origin in vibronic coupling and this work strongly supports the possibility of extending these simulations to account for those processes.\\

In conclusion, we have presented a novel computational method able to accurately describe the dynamics of the ground state magnetization of open-shell systems under the effect of vibronic coupling up to two-phonon relaxation processes. Our method generalizes previous approaches based on the effective spin Hamiltonian and shows the importance of including the effect of electronic excited states. Moving forward, this work represents a pivotal point toward delivering a complete and accurate picture of spin dynamics from first principles.
%
%

\vspace{0.2cm}
\noindent
\textbf{Acknowledgements and Funding}\\
This project has received funding from the European Research Council (ERC) under the European Union’s Horizon 2020 research and innovation programme (grant agreement No. [948493]). Computational resources were provided by the Trinity College Research IT and the Irish Centre for High-End Computing (ICHEC).

\end{document}